\documentclass[traditabstract,a4paper]{aa}  
\usepackage{graphicx}
\usepackage{longtable}
\usepackage{txfonts}

\usepackage{natbib}

\begin{document}

\title{Hot Jupiters and stellar magnetic activity}

    \titlerunning{Hot Jupiters and stellar activity}
    \authorrunning{A. F. Lanza}


   \author{A.~F.~Lanza}

   \offprints{A.~F.~Lanza}

   \institute{INAF-Osservatorio Astrofisico di Catania, Via S. Sofia, 78 
               -- 95123 Catania, Italy \\ 
              \email{nuccio.lanza@oact.inaf.it}    
             }

   \date{Received ... ; accepted ... }

    \abstract{Recent observations suggest that stellar magnetic activity may be influenced by the presence of a close-by giant planet. Specifically, chromospheric hot spots rotating in phase with the planet orbital motion have been observed during some seasons in a few stars harbouring hot Jupiters. The spot  leads the subplanetary point by a typical amount of $\sim 60^{\circ}-70^{\circ}$, with the extreme case of \object{$\upsilon$ And} where the angle is $\sim 170^{\circ}$. }{The interaction between the star and the planet is described considering the reconnection between the stellar coronal field and the magnetic field of the planet. Reconnection events produce energetic particles that moving along magnetic field lines impact onto the stellar chromosphere giving rise to a localized hot spot.}{A simple magnetohydrostatic model is introduced to describe the coronal magnetic field of the star connecting its surface to the orbiting planet. The field is assumed to be axisymmetric around the rotation axis of the star and its configuration is more general than a linear force-free field.}{With a suitable choice of the free parameters, the model can explain the phase differences between the hot spots and the  planets observed in \object{HD~179949}, \object{$\upsilon$ And}, \object{HD~189733}, and \object{$\tau$ Bootis}, as well as their visibility modulation on the orbital period and seasonal time scales. The possible presence of cool spots associated with the planets in $\tau$ Boo and \object{HD~192263} cannot be explained by the present model. However, we speculate about the possibility that reconnection events in the corona may influence subphotospheric dynamo action in those stars producing localized photospheric (and chromospheric) activity migrating in phase with their planets. }{}

   \maketitle


\section{Introduction}
About 300 extrasolar giant planets are presently 
known\footnote{See a web catalogue at: http://exoplanet.eu/}, among which $\sim 25$ percent have a
projected orbital semi-major axis  lower than 0.1 AU. Such planets are expected 
to interact significantly with their host stars, not only through tides, but possibly also
with other mechanisms. Recent investigations by \citet{Shkolniketal05,Shkolniketal08} show that 
HD~179949 and \object{$\upsilon $~Andromedae}  have chromospheric hot spots  that rotate 
with the orbital period of their inner planets. The spots are not located at 
the subplanetary point, but lead the
planet by $\sim 70^{\circ}$ in the case of \object{HD~179949} and by $\sim 170^{\circ}$ in the case of \object{$\upsilon $~And}, 
respectively. They are quite 
persistent features with lifetimes of the order of $ 350-400$ days, although they are not detected in all observing seasons. Two other stars, namely 
\object{HD 189733} and \object{$\tau$ Bootis}, show some evidence of  an excess of chromospheric variability, probably due to flaring, that is modulated with the orbital periods of their respective planets.
Again,  their active regions lead the subplanetary longitudes by $\sim 70^{\circ}$.    
Wide-band optical photometry by MOST \citep[the Microvariability and Oscillations of STar satellite; see, e.g., ][]{Walkeretal03} has given further support to the presence of an active region on $\tau$ Boo leading the planet by $\sim 70^{\circ}$. In 2004 it resembled a dark spot producing a light dip of about $0.001$ mag, while in 2005 it varied between dark and bright \citep{Walkeretal08}. 

Another intriguing observational evidence concerns \object{HD~192263}, a K dwarf for which \citet{Santosetal03}
convincingly demonstrated the presence of a close-by planet with an orbital period
of 24.35 days. Intermediate-band optical photometry by \citet{Henryetal02} showed that the star, during two
observing seasons, presented a photospheric starspot that rotated with the  orbital period of the planet
for at least three orbital cycles. The spot was not
located at the subplanetary longitude, but the planet was leading the
spot by about $90^{\circ}$. 
 
In all the above cases, an interpretation based on tidal effects can be
ruled out because there is only one signature per orbital cycle and not two. 
Following the original suggestions by \citet{Cuntzetal00} and \citet{RubenScha00}, the chromospheric
hot spot may be interpreted as an effect of the energy released by the reconnection 
 between the stellar
coronal magnetic field and the magnetic field of the planet that moves inside the star's Alfven radius 
\citep[cf. ][]{Ipetal04}. To explain the phase lag between the planet and the hot spot, 
\citet{McIvoretal06} determined the
locations at which magnetic field lines that connect the planet to the star reach the stellar surface. 
At those locations, electrons accelerated at the reconnection site hit the chromosphere
of the star producing the hot spot. Their model can account for the $\sim 70^{\circ}$ phase lag 
observed in \object{HD 179949} if the closed corona of the star extends out to the radius of the planet and the stellar
large scale dipole field is significantly tilted
 with respect to the stellar rotation axis. However, the model cannot account for
the $\sim 170^{\circ}$ phase lag observed in $\upsilon$ And by any choice of the free parameters. 

A step towards a more realistic modelling of the star-planet magnetic interaction was made by 
\citet{CranmerSaar07}, who extrapolated observed maps of the photospheric magnetic field of the Sun to compute the coronal field at different phases of activity cycle 22. The potential field source surface method was applied to derive the field assumed to become open at a distance of 2.5 solar radii from the centre of the Sun. By tracing magnetic flux tubes from the planet to the surface of the star, they were able to model the visibility of the hot spot along stellar rotation and the orbital period of the planet. The cyclic variations of the  solar field accounts for the long-term variations observed in the hot spot, while the inferred phase shifts between light curve maximum and  planetary meridian passage range between -0.2 and 0.2.

\citet{Preusseetal06}
considered a different model in which the hot spot is due to the dissipation of Alfven waves produced by the 
planet moving inside the outer stellar corona. The key difference is that Alfven waves propagate along 
characteristics that do not coincide with the magnetic field lines and that can intersect the stellar surface
virtually at any angle with respect to the direction of the planet, depending on the choice of the 
model free parameters. Therefore,
their model is capable of accounting for both the phase lags observed in \object{HD 179949} and 
\object{$\upsilon$ And}. 
However, the chromospheric flux in a hot spot can be considered comparable to that 
of a typical solar active region, i.e,
$\sim 2 \times 10^{7}$ erg cm$^{2}$ s$^{-1}$ \citep{Priest82}, which, in the case of isotropic emission, 
implies a very large Alfven wave flux at the source,  located close to the planet. In order to
have an energy flux at the source compatible with the estimates from the reconnection model 
by \citet{Ipetal04}, a highly collimated emission of Alfven waves is required, which is difficult to justify without a specific physical mechanism. 
Another difficulty of the model by \citet{Preusseetal06} is that it cannot explain the 
observations of  \object{$\tau$ Boo} and \object{HD~192263} because Alfven waves
cannot produce a cooling of the photosphere leading to the formation of dark spots.
In view of such  difficulties, in the present work we move along the way indicated by \citet{McIvoretal06}, investigating other geometries for the coronal field lines connecting the star with the planet and discuss the possible effects of coronal field reconnection on the dynamo action occurring in the convection zone of the star. 
 
\begin{table*}
\caption{Stellar and planetary parameters for different cases of star-planet magnetic interaction}
\begin{tabular}{cccccccc}
\hline
Star & Sp. type & $P_{\rm rot}$ & $R$ & $A_{0}$  & $P_{\rm orb} $ & $u_{\rm A}/u_{\rm orb}$ & Ref.$^{a}$\\
 & & (d) & (R$_{\odot}$) & (AU) & (d) & \\
\hline
 & & & &  & \\
HD~179949      & F8V & $ \sim 7$ & 1.25 &  0.045 & 3.09 & 14.6 & S08\\
$\upsilon$ And & F7V & $\sim 12$ & 1.25 & 0.059 & 4.62 & 16.7 & S08  \\
$\tau$ Boo     & F7IV & 3.2 & 1.62 & 0.046 & 3.31 & 15.3 & S08, D08 \\
HD~189733      & K1V & 11.7 & 0.80 & 0.031 & 2.22 & 15.2 & S08 \\
HD~192263      & K2V & $\approx 24$ & 0.78 &  0.15  & 24.35 & 34.6 & S03 \\ 
& & & &  \\
\hline
\label{table_data}
\end{tabular}
~\\
$^{a}$ References: S08: Shkolnik et al. (2008); D08: Donati et al. (2008); S03: Santos et al. (2003).
\end{table*}

\section{Model of the stellar coronal field}
\label{chromo_model}

We  focus on the interpretation of chromospheric spots moving synchronously with  hot Jupiters. 
{ In Table~\ref{table_data}, we list the relevant stellar and planetary parameters, i.e., from the first to the eighth column, respectively, the name of the star, its spectral type, rotation period $P_{\rm rot}$, radius $R$,  semi-major axis of the planetary orbit (assumed to be circular) $A_{0}$, orbital period $P_{\rm orb}$, ratio of the Alfven velocity to the planet orbital velocity $u_{\rm A}/u_{\rm orb}$, calculated for a magnetic field of 0.15 G and a coronal density at the orbital distance of the planet of $2 \times 10^{4}$ protons cm$^{-3}$; and the references. }  
We assume that magnetic reconnection at the intersection between stellar and planetary magnetospheres accelerates electrons to supra-thermal energies. A fraction of those high-energy particles  travels inward along magnetic field lines and  
impacts onto the dense chromospheric layers of the star, producing a localized heating which gives rise to the hot spot. 

The position and the extension of the spot depend on the geometry of the stellar coronal field that conveys energetic particles onto the chromosphere. We want to show that a coronal field geometry actually exists that can explain the observed phenomenology, although we shall not investigate how it can be produced and maintained.  

We assume that the orbital plane of the planet coincides with the stellar equatorial plane and that its orbit is circular. We consider a spherical polar reference frame with the origin at the centre of the star, the $\hat{z}$ axis along the stellar rotation axis and assume the radius $R$ of the star as the unit of measure. We indicate  the radial distance from the origin with $r$, the colatitude measured from the North pole with $\theta$, and the azimuthal angle with $\phi$. 

In the low-density environment of the stellar corona, the Alfven velocity is at least one order of magnitude larger than the orbital velocity of the planet \citep[cf. e.g., ][ and Table~\ref{table_data}]{Ipetal04}, therefore we can assume that the corona is in a state of magnetostatic equilibrium. For the sake of simplicity, we consider a specific class of magnetostatic equilibria, i.e., that investigated by \citet{Neukirch95} which includes as a particular case the linear force-free equilibria in spherical geometry previously described by \citet{Chandrasekhar56} and \citet{ChandrasekharKendall57}. 
{
Generalizing a previous approach by Low, Neukirch assumes that the current density in the corona has two components, one directed along the magnetic field, as in force-free models, and the other flowing in a direction perpendicular to the local gravitational acceleration.
}
Note that the gravitational field of the star dominates over the centrifugal force and the gravitational field of the planet over most of the stellar corona, because the star is rotating quite slowly 
\citep[the rotation period of \object{HD 179949} is about 7 days, while that of \object{$\upsilon$ And} is about 12 days; cf. ][]{Shkolniketal05,Shkolniketal08} and the mass of the planet is of the order of 
$10^{-3}$ stellar masses. Therefore, the spherical equilibria computed by \citet{Neukirch95} under the assumption of a purely radial stellar gravitational field are good approximations for our case. 

{ The  plasma pressure, density and temperature in the corona are not derived consistently by solving the energy equation, rather they are determined a posteriori  to verify the magnetohydrostatic balance and the equation of state. Nevertheless, in view of our ignorance of the plasma heating function, this approach is allowable to investigate analytically coronal field geometries of a wider class that  linear force-free fields.}

Non-axisymmetric coronal fields give rise to an additional modulation of the  hot spot intensity with the rotation period of the star because magnetic field lines are rooted into the star.  Therefore, we shall consider only axisymmetric configurations. Since the orbital radius of the planet is significantly greater than the stellar radius, the field component with the slowest decay as a function of the radial distance (i.e., the dipole-like component) is the one leading to the strongest interaction. For simplicity, we shall consider only that component assuming that it dominates over all the other components. This can be justified in the case of a solar-like $\alpha$-$\Omega$ dynamo, as the axisymmetric dynamo mode with dipole-like symmetry with respect to the equator is the one with the strongest excitation rate and dominates the global field geometry outside the star.   

{
Under these hypotheses, we consider the magnetic field configuration indicated as case II in Sect. 4 of \citet{Neukirch95},  whose field components in the axisymmetric case ($l=1$, $m=0$) are: 
\begin{eqnarray}
B_{r} & = & 2B_{0} \frac{R^{2}}{r^{2}} g(q) \cos \theta, \nonumber \\
\label{field_conf}
B_{\theta} & = & - B_{0} \frac{R^{2}}{r}\frac{d}{dr} g(q)  \sin \theta, \\
B_{\phi} & = & \alpha B_{0} \frac{R^{2}}{r} g(q)  \sin \theta \nonumber,  
\end{eqnarray}
where $2B_{0}$ is the surface magnetic field at the North pole of the star,  $\alpha$ the helicity parameter of the corresponding linear force-free field, and the function $g$ is defined as:
\begin{equation}
g (q) \equiv \frac{[b_{0} J_{-3/2}(q) + c_{0} J_{3/2}(q)]\sqrt{q}}{[b_{0} J_{-3/2}(q_{0}) + c_{0} J_{3/2}(q_{0})] \sqrt{q_{0}}},
\end{equation}
where $b_{0}$ and $c_{0}$ are free constants, $J_{-3/2}$ and $J_{3/2}$ are Bessel functions of the first kind of order $-3/2$ and $3/2$, respectively, $q \equiv |\alpha | (r +a)$ and $q_{0} \equiv |\alpha | (R +a)$, with $a$ a parameter measuring the deviation from the corresponding linear force-free field. The function $g$ is a linear combination of the two independent solutions $\sqrt{r+a}J_{3/2}(q)$ and $\sqrt{r+a} N_{3/2}(q) = \sqrt{r+a} J_{-3/2}(q)$ obtained by Neukirch for his case II 
\citep[see Eqs. (40)-(43) in ][]{Neukirch95}\footnote{For the relationship between the Neumann functions 
$N_{p}$ and the Bessel functions of the first kind $J_{p}$, 
see, e.g., \citet{Smirnov64}.}. 
When $a=0$ we recover the linear axisymmetric force-free solution of \citet{Chandrasekhar56} and \citet{ChandrasekharKendall57} as given in, e.g., \citet{Flyeretal04}. The physical meaning of the parameter $a$ can be understood by noting that the transformation $r+a \rightarrow r$ changes our field into a linear force-free field. Therefore, our field is a somewhat radially "squeezed" version of a linear force-free field, with the additional confining forces being the gravity and the pressure gradient and 
$a$ measuring the amount of radial compression. 
}
We introduce the absolute value of $\alpha$ in the definition of the radial variable $q$, so Eq.~(\ref{field_conf}) is valid for both positive and negative $\alpha$. Changing the sign of $\alpha$ is equivalent to change the sign of  $B_{\phi}$ leaving $B_{r}$ and $B_{\theta}$ unaltered, 
as can be deduced by comparing Eqs. (18) and (24) of \citet{Neukirch95}. 

The magnetic field given by Eqs.~(\ref{field_conf}) does not extend to the infinity, but it is confined to the region between $r=R$ and $r=r_{\rm L}$, where $r_{\rm L}$ corresponds to the first zero of $g$ \citep[see ][ for a discussion of the boundary conditions at $r=r_{\rm L}$]{Chandrasekhar56}. Therefore, it can be used to model the inner, closed corona of the star, but not the outer region where the field lines become radial and the stellar wind is accelerated. \citet{Schrijveretal03} estimate that the extent of the closed corona is $r_{\rm L} \approx 2.5$ for the Sun and  increases remarkably for more active stars, e.g, $r_{\rm L} \approx 19$ for a star having a surface magnetic flux density or a coronal X-ray flux 10 times greater than the Sun. Since the stars we are considering rotate faster and generally have higher X-ray fluxes than the Sun 
\citep[cf., e.g., ][]{Shkolniketal05,Shkolniketal08,Saaretal07}, 
we estimate that our model can be extended up to a radial distance of $\approx 20-30$ before becoming inadequate (cf. Sect.~\ref{application1}).

{
To derive the path of the magnetic field lines, we use the approach of \citet{Flyeretal04}, introducing a 
flux function:
\begin{equation}
A(r, \theta) \equiv B_{0} R^{2} g (q) \sin^{2} \theta, 
\label{theta_vs_r}
\end{equation}
so that the magnetic field is given by: 
${\vec B} = \frac{1}{r \sin \theta} \left[ \frac{1}{r} \frac{\partial A}{\partial \theta} \hat{\vec r} - \frac{\partial A}{\partial r} \hat{\vec \theta} + \alpha A \hat{\vec \phi} \right]$. Magnetic field lines lie  over surfaces of constant $A(r, \theta)$, 
as can be deduced by noting that ${\vec B} \cdot \nabla A = 0$. The variation of the azimuthal angle along a given field line can be derived from the equation: 
\begin{equation}
\frac{dr}{B_{r}}  =  \frac{r \sin \theta d \phi}{B_{\phi}},    
\end{equation}
where ${\vec dl} \equiv (dr, r d\theta, r \sin \theta d \phi)$ is the line element. 
Substituting the expressions of the magnetic field components, it gives:
\begin{equation}
\phi(q) - \phi(q_{0}) = \frac{1}{2} \int_{q_{0}}^{q} \frac{d q^{\prime}}{\cos \theta (q^{\prime})}. 
\end{equation}
Making use of Eq.~(\ref{theta_vs_r}), and indicating by $q_{e}$  the value of $q$ at which the field line reaches the equatorial plane, i.e., 
$g(q_{e})= \sin^{2} \theta_{0}$, where $\theta_{0}$ is the initial colatitude of the field line on the stellar surface, we find:
\begin{equation}
\phi(q) - \phi(q_{0}) = \frac{1}{2} \int_{q_{0}}^{q} \sqrt{ \frac{g(q^{\prime})}{g(q^{\prime}) 
- g(q_{e}) } } d q^{\prime}. 
\label{delta_phi}
\end{equation}
Eq.~(\ref{delta_phi}) is valid for a magnetic field line extending to the equatorial plane because the integral exists all along  the closed interval $[q_{0}, q_{e}]$,
as it is shown in Appendix~\ref{appendixB}.
}

\section{Applications}
\label{application1}

{
The magnetic configuration connecting the chromospheric spot to the planet with the observed phase shift is not unique in the framework of our model, even in the force-free case. 
A complete exploration of the parameter space would be long and complex, so we limit ourselves to present some possible configurations obtained by fixing $b_{0}=-1.1$, $c_{0}=1.0$, and
$a=1.4$, and varying $\alpha$  and $\theta_{0}$  until the magnetic field lines reach the equatorial plane at the planet orbital radius $r_{\rm e}$  with an azimuthal angle difference $\Delta \phi_{\rm e}$ compatible with the observations. 
 Our choice of $b_{0}$ and $c_{0}$ is motivated by the need of having coronal field lines with an azimuthal angle $\Delta \phi_{\rm e}$ sufficient  to account for the large phase gap observed in $\upsilon$ And. By making some experiments with different values of those parameters, we selected a  couple of values out of virtually infinite possible choices, only for illustrative purposes.
In Table~\ref{table_conf}, for each star listed in the first column, we report the values of $a$, $\alpha$, $\theta_{0}$, $r_{\rm e}$, $\Delta \phi_{\rm e}$, and 
 $r_{\rm L}$ in the columns from the second to the seventh, respectively. Note that 
$r_{\rm L}$ corresponds to the first zero of $g$ at $q_{\rm L}= 3.609$. For \object{HD~179949} and \object{$\upsilon$ And} we also list the parameters for two force-free models.} 

\begin{figure}[t]
\includegraphics[width=8cm,height=12cm]{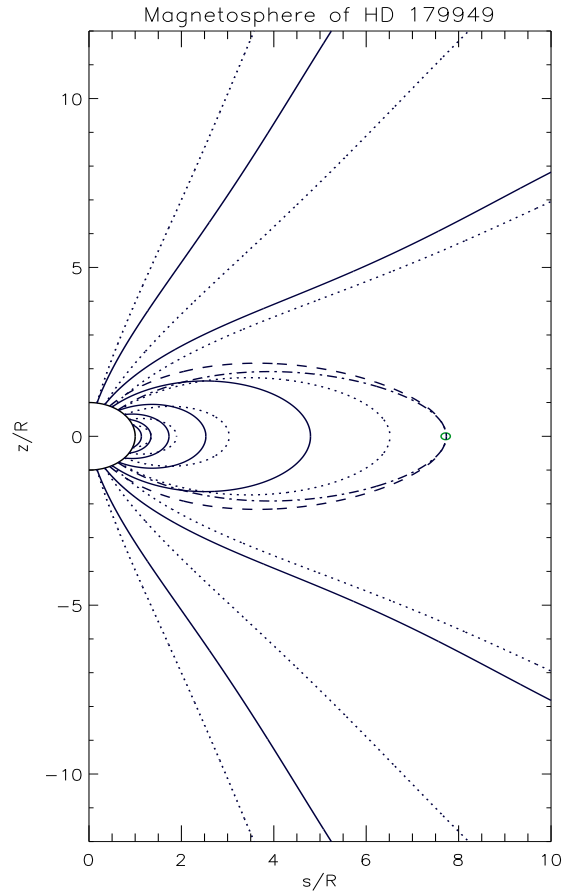} 
\caption{Projections of the magnetic field lines of the magnetosphere of HD~179949 
on a meridional plane. Configurations are displayed 
for Neukirch's models with two sets of parameters, i.e., $\alpha= -0.10$, $a=1.4$ (solid lines) and 
$\alpha=-0.12$ and $a=0$ (dotted lines). The distance $z$ from the equatorial plane and the distance $s = r \sin \theta$ from the rotation axis are measured in units of the stellar radius 
$R$. The small green circle indicates the planet assumed to have a radius of 0.1. The dot-dashed line indicates a field line with initial colatitude $38.^{\circ}85$ connecting the surface of the star with the planet for the model with $\alpha=-0.10$ and $a=1.4$. The dashed line is the same for the force-free model ($\alpha=-0.12$, $a=0.0$) and has an initial colatitude of  $25.^{\circ}62$. }
\label{mag_hd179949}
\end{figure}

The best observed example of star-planet interaction is provided by \object{HD~179949}, so we describe its models in some detail.  
A meridional section of its magnetosphere  computed for
$\alpha= -0.10$ and $a=1.4$ is shown in Fig.~\ref{mag_hd179949}. The
field lines at an initial colatitude $\theta_{0} = 38.^{\circ}85$  reach the planet on the equatorial plane at $r_{\rm e}=7.72$ with $\Delta \phi_{\rm e}$ comparable with the observations 
 -- note that typical uncertainties on the observed phase lags are $\pm (15^{\circ}-20^{\circ})$. 
The corresponding force-free field with $\alpha=-0.10$ gives $\theta_{0}=25.^{\circ}0$ and 
$\Delta \phi_{e} = 53.^{\circ}.7$ which is too small to explain the observations. Therefore, instead of changing $b_{0}$ or $c_{0}$, we decided to vary $\alpha$ and find an acceptable solution 
 with $\alpha=-0.12$ that is also shown in Fig.~\ref{mag_hd179949}.  In this case, the field lines reaching the planet have an initial colatitude of $\theta_{0} = 26.^{\circ}62$. The angle $\Delta \phi$ 
between the footpoint on the stellar surface and the point  at distance
$r$ along a field line connecting to the planet is plotted vs. the radial distance 
in Fig.~\ref{phi_hd179949}. At the orbital radius of the planet this corresponds to an angle of $-66.^{\circ}4$ in the case of the field lines with $a=1.4, \alpha=-0.10$, giving a phase lag of $\sim 0.18$ between the planet and the hot spot, or to an angle of $-73.^{\circ}8$
in the case of the force-free field lines with $\alpha=-0.12$, giving a phase lag of $\sim 0.205$. The latitudinal extension of the hot spot on the surface of the star depends on the radius of the planetary magnetosphere, i.e., the distance from the planet at which magnetic reconnection occurs.
If we assume a magnetospheric radius $R_{\rm m} = 4 R_{\rm p}$, where $R_{\rm p} = 0.1$ is the radius of the planet, then the radial boundaries of the magnetic reconnection region are 
at $7.32$ and $8.12$, respectively. The colatitudes on the stellar surface of the magnetic field lines reaching the equatorial plane at those distances are $39.^{\circ}17$ and $38.^{\circ}60$,
respectively. On the other hand, the longitudinal extension of the region is approximately 
$6^{\circ}$, as derived by the intersections of the surface of the star with  the projections on the equatorial plane of the  field lines tangent to the planet's magnetosphere.   Therefore, the energy carried by high-energy electrons is focussed onto a quite small region of the stellar chromosphere, producing a remarkable contrast with respect to the background emission. 

The hot spot on \object{HD~179949} stays in view for about 0.5 of an orbital cycle \citep[see Fig. 6 in ][]{Shkolniketal08}, which suggests that the inclination of the stellar rotation axis on the line of sight is quite high and the latitude of the spot  quite low. Therefore, the magnetic 
configuration with $a=1.4$ is favoured with respect to the force-free configuration because it gives a spot at a lower latitude ($\sim 50^{\circ}$ vs. $\sim 65^{\circ}$). 
{ However, the non-force-free magnetic configuration is associated with significant perturbations of the thermal equilibrium of the stellar corona. Their relative amplitudes depend not only on the intensity and geometry of the field, but also on the background pressure and density distributions 
which are largely unknown \citep[cf. Eqs. (28) and (30) of ][]{Neukirch95}. An illustrative example is shown in Fig.~\ref{corona_hd179949} where the relative variations of the temperature, pressure and density in a meridional plane are plotted for a background isothermal corona with a field having the geometry computed for HD~179949. The background pressure and  density are proportional to  $\exp[\Lambda (1/r -1/R)]$, with $\Lambda = 10.4 R$, and the ratio of the plasma pressure to the magnetic pressure at the base  of the corona  
$\beta_{0} \sim 5$, while the unperturbed background temperature is $1.66 \times 10^{6}$ K. The largest temperature and density perturbations are localized at high latitudes and quite close to the star. There the decrease of the plasma pressure provides a force that pushes the magnetic field  lines toward the poles and the surface of the star, thus accounting for the squeezing characteristic of the considered non-force-free configurations with respect to the corresponding force-free models with $a=0$, as discussed above.}

If the magnetic field of the planet is a dipole aligned with the planet's rotation axis and that axis is perpendicular to the orbital plane, then the planet's field is directed in the meridional plane close to the equator and reconnection occurs mainly with the stellar $B_{\theta}$ component. If the star has a solar-like magnetic cycle characterized by cyclic inversions of the sign of 
$B_{\theta}$, then  reconnection delivers most of the energy when $B_{\theta}$ is opposite to the planet's field. This implies that the mean energy of the accelerated electrons is  a function 
of the cycle phase of the star, with a significantly lower value when the two field components have the same sign. This may explain the on/off character of the star-planet magnetic interaction reported by \citet{Shkolniketal08}. Recent observations of the evolution of the surface magnetic field in $\tau$ Boo support the hypothesis that magnetic cycles in F-type dwarfs may be significantly shorter than the solar cycle \citep{Donatietal08}.

Models similar to that presented for \object{HD~179949} can be computed for \object{$\tau$ Boo} and 
\object{HD~189733} for which the hot spots lead the respective planets by $\sim 60^{\circ}-80^{\circ}$. 
{
On the other hand,   
in the case of \object{$\upsilon$ And}, we need a higher value of $\alpha$ to account for the larger phase shift. In Fig.~\ref{mag_nu_And}, we show meridional sections of the magnetosphere of the star  computed for a model with  $a=1.4$ and a force-free model, respectively. Note the lower latitudes of the footpoints of the field lines connecting to the planet in comparison to the case of HD~179949. 
In Fig.~\ref{phi_nu_And}, we plot the variation of the azimuthal angle along magnetic field lines for
the models shown in Fig.~\ref{mag_nu_And}. 

In the case of \object{HD~192263}, we may assume that chromospheric plages be associated with the photospheric spots lagging the planet by  $\approx 90^{\circ}$. The former can be explained by the present model by assuming a positive $\alpha$, whereas in all the other cases the sign of $\alpha$ is negative (cf. Table~\ref{table_conf}).
Magnetic configurations with opposite values of $\alpha$ have the same energy and opposite magnetic helicity, so there is no preference for a particular sign of $\alpha$ from the point of view of the energy balance or field stability. Note that a linear force-free configuration with a given value of 
$\alpha \la 3$ is stable  provided that its magnetic helicity exceeds a certain value \citep[cf., e.g., ][]{Breger85}. 

We suggest that the sign of $\alpha$ may be related with the conservation of angular momentum in radial plasma motions along field lines. In other words, if there is a mass flow along the field lines, it tends to bend the lines toward negative $\Delta \phi$ (i.e., $\alpha < 0$) when  rotation is in the same sense of the orbital motion of the planet, or toward positive $\Delta \phi$ (i.e., $\alpha > 0$) when it is in the opposite sense. 
The selection of the sign of $\alpha$ takes place close to the stellar surface since plasma motions cannot affect the field geometry in the low-$\beta$ environment of the outer corona.  
This effect cannot be described in detail in the framework  our magnetohydrostatic approach, but it could be incorporated in more general models.  

It is important to note that in the case of HD~192263 the absolute value of $\alpha$ is $4-5$ time smaller than in the other cases, because the relative orbital radius of the planet is correspondingly larger. As a consequence, the relative radius of the  closed corona turns out to be $4-5$ times larger than in the other cases. However, it can be significantly reduced by making a different choice 
for the free parameters $b_{0}$ and $c_{0}$ in the definition of the function $g$, in order to have its first zero at a value smaller than $q_{\rm L}=3.609$. 
}

\begin{figure}[t]
\includegraphics[width=8cm,height=6cm]{09753.f2} 
\caption{Variation of the azimuthal angle $\Delta \phi$ along a magnetic field line starting at the  surface of HD~179949 as a function of the radial distance in units of the stellar radius. The solid line is for the Neukirch's model with $\alpha=-0.10$, $a=1.4$, and an initial colatitude  $\theta_{0} = 38^{\circ}.85$; the dotted line is for $\alpha=-0.12$, $a=0$, and 
$\theta_{0} = 25^{\circ}.62$. The vertical dot-dashed line indicates the orbital radius of the planet of HD~179949.}
\label{phi_hd179949}
\end{figure}

\begin{figure*}[t]
\includegraphics[width=8cm,height=16cm,angle=90]{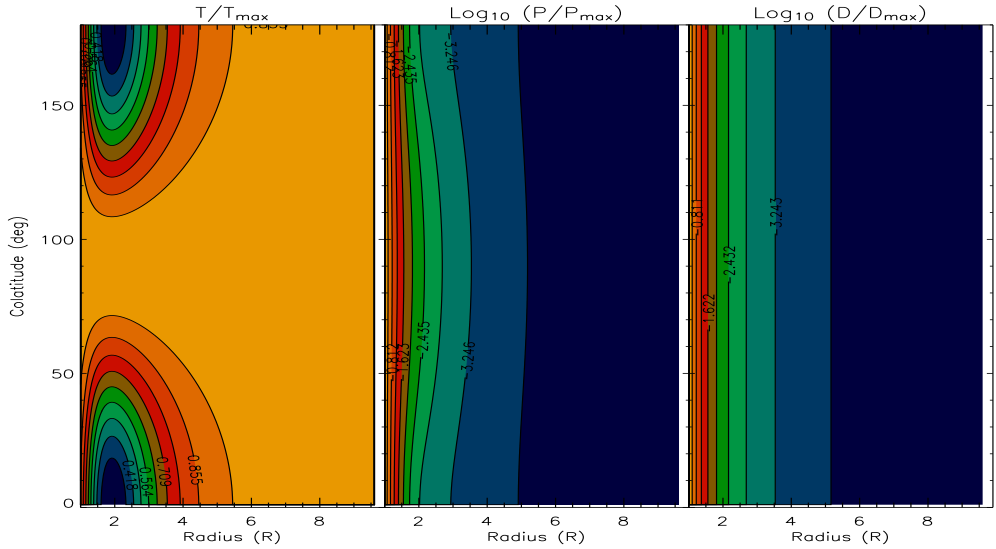} 
\caption{Relative temperature (left panel), logarithm of the relative pressure (middle panel) and of the relative density (right panel) for an illustrative coronal model computed for HD~179949 (see  text for details). }
\label{corona_hd179949}
\end{figure*}
\begin{figure}[t]
\includegraphics[width=8cm,height=12cm]{09753.f4} 
\caption{Projections of the magnetic field lines of the magnetosphere of $\upsilon$ And
on a meridional plane. Configurations are displayed 
for Neukirch's models with two sets of parameters, i.e., $\alpha= -0.18$, $a=1.4$ (solid lines) and 
$\alpha=-0.208$ and $a=0$ (dotted lines). The distance $z$ from the equatorial plane and the distance $s = r \sin \theta$ from the rotation axis are measured in units of the stellar radius 
$R$. The small green circle indicates the planet assumed to have a radius of 0.1. The dot-dashed line indicates a field line with initial colatitude $48.^{\circ}90$ connecting the surface of the star with the planet for the model with $\alpha=-0.18$ and $a=1.4$. The dashed line is the same for the force-free model ($\alpha=-0.208$, $a=0.0$) and has an initial colatitude of  $32.^{\circ}85$. }
\label{mag_nu_And}
\end{figure}
\begin{figure}[t]
\includegraphics[width=8cm,height=6cm]{09753.f5} 
\caption{Variation of the azimuthal angle $\Delta \phi$ along a magnetic field line starting at the  surface of $\upsilon$ And as a function of the radial distance in units of the stellar radius. The solid line is for a Neukirch's model with $\alpha=-0.18$, $a=1.4$, and an initial colatitude
$\theta_{0} = 48^{\circ}.90$; the dotted line is for $\alpha=-0.208$, $a=0$, and 
$\theta_{0} = 32^{\circ}.85$. The vertical dot-dashed line indicates the orbital radius of the planet 
of $\upsilon$ And.}
\label{phi_nu_And}
\end{figure}

\begin{table}
\caption{Model parameters for different cases of star-planet magnetic interaction}
\begin{tabular}{ccccccc}
\hline
Star & $a$ & $\alpha$ & $\theta_{0}$ & $r_{\rm e}$ & $\Delta \phi_{\rm e}$ & $r_{\rm L}$ \\
 & & & (deg) & ($R$) & (deg) & ($R$) \\
\hline
 & & & & & & \\
HD~179949      & 1.4 & -0.100 & 38.85 & 7.72 & 66.4 & 34.7 \\
               & 0.0 & -0.120 & 26.62 & 7.72 & 73.8 & 30.1 \\
$\upsilon$ And & 1.4 & -0.180 & 48.90 & 10.16 & 173.6 & 18.6 \\
               & 0.0 & -0.208 & 32.85 & 10.16 & 173.7 & 17.3 \\
$\tau$ Boo     & 1.4 & -0.130 & 44.50 & 6.12 & 72.7 & 26.3 \\
HD~189733      & 1.4 & -0.100 & 38.48 & 8.32 & 77.9 & 34.7 \\
HD~192263      & 1.4 &  0.025 & 18.24 & 41.19 & -106.7 & 143.0 \\ 
& & & & & \\
\hline
\label{table_conf}
\end{tabular}
\end{table}
\section{Discussion}

 The model presented in Sect.~\ref{chromo_model} can explain phase gaps up to 
$\sim 180^{\circ}$, which is not possible in the framework of the previous model by \citet{McIvoretal06}. Our model can accounts for chromospheric hot spots, but it
cannot account for cool dark spots moving in phase with the planet, as suggested by MOST  observations of $\tau$ Boo in 2004 or in the case of \object{HD~192263}. Therefore, we speculate about the possibility that the magnetic reconnection in the stellar corona may perturb some process responsible for the dynamo action inside the star. Specifically, the effect leading to regeneration of the poloidal field from the toroidal field in a solar-like dynamo, i.e., the 
$\alpha_{\rm D}$-effect\footnote{Note that the parameter $\alpha_{\rm D}$ of the dynamo theory, considered in this Section, in Sect.~\ref{conclusions}, and in Appendix~\ref{appendixa}, is conceptually different from the parameter $\alpha$ characterizing force-free coronal field in Sects.~\ref{chromo_model} and 
\ref{application1}. Therefore, we decided to mark it with the index D to avoid any confusion.}, may be perturbed by magnetic helicity losses  at the reconnection sites in the corona \citep{BlackmanBranden03, BrandenSubra05}. In Appendix~\ref{appendixa} we propose some preliminary ideas about a mechanism that may couple coronal reconnection 
with the $\alpha_{\rm D}$-effect  in the convective layers immediately below the surface. 
 If a significant dynamo action takes place in those subsurface layers, as suggested by
\citet{Brandenburg05,Brandenburg07}, this may give rise to a local amplification of the magnetic field that will manifest itself with localized photospheric (and chromospheric) activity, 
 including cool spots. 
Models with a longitude-depend $\alpha_{\rm D}$ effect have been explored by, e.g., \citet{Mossetal02} and show that the maximum  magnetic field strength  usually falls within $\sim 15^{\circ}-20^{\circ}$ from the maximum of the $\alpha_{\rm D}$ effect. Differential rotation tends to wipe out non-axisymmetric components of the magnetic field, but active longitudes may survive at latitudes of minimal shear as discussed by, e.g., \citet{BigazziRuzmaikin04}. 

If this scenario is correct, we expect to find not only hot chromospheric spots but also cool photospheric spots at the longitudes where magnetic field lines connecting the planet with the star intersect the stellar surface. The phase lags observed in \object{$\tau$ Boo} and 
\object{HD~192263} may be explained in the framework of the coronal field model described in Sect.~\ref{application1}. 

{ In the case of the Sun, it is well known that new active regions have a marked preference to form close to existing active regions giving rise to complexes of activity with lifetimes up to $5-6$ months 
\citep[e.g., ][]{HarveyZwaan93}. We  conjecture that a possible cause of this increased probability of new magnetic flux  emergence may be a localized enhancement of the subphotospheric dynamo action promoted by helicity losses in the coronal part of an already existing active region. However, the correlation between the rate of coronal mass ejection and the rate of flux emergence in a complex of activity needs to be studied in detail to test this conjecture.}

\section{Conclusions}
\label{conclusions}

We extended the model proposed by \citet{McIvoretal06} considering a non-potential magnetic field configuration for the closed corona of a star accompanied by a hot Jupiter.  We showed that a tilt of the axis of symmetry of the magnetic field with respect to the rotation axis of the star is not needed to reproduce the observed phase lags between the hot spots and the planets in \object{HD~179949}, 
\object{$\upsilon$ And}, \object{$\tau$ Boo}, and \object{HD~189733}. This agrees with  mean-field dynamo models that predict that the large-scale stellar dipole-like field is aligned with the rotation axis when  differential rotation is significant, as recently found in the case of $\tau$ Boo \citep{Catalaetal07,Donatietal08}. The novelty of the present model is that it can easily  account for a phase gap as large as that observed in $\upsilon$ And, which is not possible with previous approaches \citep{McIvoretal06,CranmerSaar07}.

We also found that coronal linear force-free fields tend to give hot spots at quite high latitudes, making it difficult to explain the observed modulation of their visibility. Non-force free-equilibria, such as those computed by \citet{Neukirch95}, give a significantly lower spot latitude  for an appropriate choice of the parameter $a$, allowing us to fit  the observations without assuming a high inclination for the stellar rotation axis and the orbital plane of the planet. 

The on/off nature of the star-planet interaction suggested by \citet{Shkolniketal08} can also be explained in the framework of our model as a consequence of stellar activity cycles 
{ \citep[see  ][ for a model based on observed solar cycle variations of the magnetic field]{CranmerSaar07}}. 

We speculate about the role of magnetic helicity losses due to reconnection events in the stellar corona induced by the planet and suggest that this may give rise to a longitudinal dependent $\alpha_{\rm D}$-effect in the stellar dynamo process. 
Note that the mechanism we suggest may provide a physical basis for the $\alpha_{\rm D}$-effect perturbation model originally proposed by \citet{Cuntzetal00}. Therefore, it is worthwhile to investigate dynamo models with an $\alpha_{\rm D}$-effect varying in phase with the longitude of the planet to see whether they can reproduce the observations. This kind of models could provide an explanation for a modulation of the photospheric activity in phase with the planet that cannot be interpreted by means of the chromospheric heating model. 

Zeeman Doppler Imaging (ZDI) techniques are  of great interest to map the magnetic fields in the photospheres of planet-harbouring stars, thus providing  boundary conditions for an extrapolation to their coronal fields \citep{Moutouetal07,Donatietal08}. Although non-potential configurations for assigned boundary conditions may  not be unique, including ZDI information in our approach may give in principle the possibility of a self-consistent model of the stellar coronal field to study its interaction with the planet's magnetosphere \citep[cf. e.g., ][]{Wiegelmannetal07}.

\begin{acknowledgements}
The author wishes to thank an anonymous Referee for a careful reading of the manuscript and valuable comments that greatly helped to improve the present work. 
AFL is grateful to Drs. P.~Barge, C.~Moutou and I.~Pagano for drawing his attention to the interesting problem of star-planet magnetic interaction. This work has been partially supported by  the Italian Space Agency (ASI) under contract  ASI/INAF I/015/07/0,
work package 3170. Active star research and exoplanetary studies at INAF-Catania Astrophysical Observatory and the Department of Physics
and Astronomy of Catania University is funded by MUR ({\it Ministero dell'Universit\`a e Ricerca}), and by {\it Regione Siciliana}, whose financial support is gratefully
acknowledged. 
This research has made use of the ADS-CDS databases, operated at the CDS, Strasbourg, France.
\end{acknowledgements}

\appendix

\section{On the integrability of the function in Eq.~(\ref{delta_phi})}
\label{appendixB}
{
We prove that the  integral appearing in:  
\begin{equation}
\phi(q_{e}) - \phi(q_{0})  =  \frac{1}{2} \int_{q_{0}}^{q_{e}} \sqrt{ \frac{g(q^{\prime})}{g(q^{\prime}) 
- g(q_{e}) } } d q^{\prime}. 
\end{equation}
is finite, in spite of the denominator of the integrand vanishing at $q=q_{e}$. Let us select a point $q_{1}=q_{e} -\epsilon$, where $\epsilon $ is a small positive number. Applying Lagrange's formula, we can write the denominator of the integrand as:  $g(q)-g(q_{e})=g^{\prime}(\xi) (q -q_{e})$, where $q_{1} \leq q \leq q_{e}$, $q \leq \xi \leq q_{e}$, and $g^{\prime}$ is the derivative of $g$. 
If $g^{\prime}$ has no zeros in the interval $[q_{1}, q_{e}]$, the function
$\sqrt{g(q)/g^{\prime}(\xi)}$ is continuous and limited in the same interval. Indicating its upper bound with $H$, we have: 
\begin{eqnarray}
\phi(q_{e}) - \phi(q_{0}) & = & \frac{1}{2} \int_{q_{0}}^{q_{e}} \sqrt{ \frac{g(q^{\prime})}{g(q^{\prime}) 
- g(q_{e}) } } d q^{\prime}  = \nonumber \\
  =  \frac{1}{2} \int_{q_{0}}^{q_{1}} \sqrt{ \frac{g(q^{\prime})}{g(q^{\prime}) 
- g(q_{e}) } } d q^{\prime} & + & \frac{1}{2} \int_{q_{1}}^{q_{e}} \sqrt{ \frac{g(q^{\prime})}{g(q^{\prime}) 
- g(q_{e}) } } d q^{\prime}  \nonumber \\
  <  \frac{1}{2} \int_{q_{0}}^{q_{1}} \sqrt{ \frac{g(q^{\prime})}{g(q^{\prime}) 
- g(q_{e}) } } d q^{\prime} & + & \frac{H}{2} \int_{q_{1}}^{q_{e}} \frac{d q^{\prime}}{\sqrt{q_{e}-q^{\prime}}} \nonumber \\
   =  \frac{1}{2} \int_{q_{0}}^{q_{1}} \sqrt{\frac{g(q^{\prime})}{g(q^{\prime})-g(q_{e})}} d q^{\prime} 
& + & H \sqrt{q_{e}-q_{1}}. 
\end{eqnarray}
This formula can  be applied to estimate an upper bound for the error made when the numerical integration is truncated at a value $q_{1} < q_{e}$. Specifically, in the case of the values listed in Table~\ref{table_conf}, we have fixed $q_{1}$ by requiring that $\cos \theta(q_{1}) = 0.01$, thus finding an error smaller than $3.^{\circ}6$ for $\Delta \phi_{\rm e}$. 
}

\section{Conjecture on a mechanism to perturb stellar dynamo action}
\label{appendixa}

The value of the parameter $\alpha_{\rm D}$ of the dynamo process depends, among others,  on the mean value of the current helicity density ${\vec j} \cdot {\vec B}$ \citep[cf. e.g., Eq. (9.58) of ][]{BrandenSubra05}.
Therefore, a perturbation of ${\vec j} \cdot {\vec B}$ in the subsurface layers may produce a corresponding perturbation of $\alpha_{\rm D}$. The current density in Neukirch's model 
\citep[cf. Eq. (7) in ][]{Neukirch95} is:
\begin{equation}
{\vec j} = \alpha {\vec B} + \nabla F \times \nabla \psi, 
\label{j_eq}
\end{equation} 
where $\psi$ is the gravitational potential and $F=F(\nabla \psi \cdot {\vec B}, \psi)$ a suitable scalar function defined in \citet{Neukirch95}. Since $\nabla F \times \nabla \psi = \nabla \times (F \nabla \psi)$ , taking the divergence of Eq.~(\ref{j_eq}) we find that  $({\vec B} \cdot \nabla) \alpha = 0$, i.e., $\alpha$ is constant along a given magnetic field line. The current helicity at the surface of the star
is given by: 
\begin{equation}
{\vec j} \cdot {\vec B} = \alpha B^{2} + {\vec B} \cdot \nabla \times (F \nabla \psi).  
\end{equation}
If the perturbations of the coronal magnetic field due to reconnection events do not directly affect the field at the surface, $B^{2}$ and ${\vec B} \cdot \nabla \times (F \nabla \psi)$ are not  modified by the motion of the planet. However, reconnection events induced by the planet will change the value of $\alpha$ in the corona and such changes will be conducted on the Alfven travel time to the stellar surface where they change ${\vec j} \cdot {\vec B}$. Since the magnetic field at the surface is mainly radial, ${\vec j} \cdot {\vec B} \simeq j_{\rm n} B_{\rm n}$, where $j_{\rm n}$ and $B_{\rm n}$ are the components of ${\vec j}$ and ${\vec B}$ normal to the surface bounding the star. The perturbation of 
$j_{\rm n} B_{\rm n}$ is then conducted into the subsurface layers because $j_{\rm n}$ and $B_{\rm n}$ are continuous across the bounding surface. 

According to this scenario, a localized subphotospheric perturbation of $\alpha_{\rm D}$ is produced, that moves in phase with the planet. A corresponding localized increase in the mean field gradient may promote a magnetic buoyancy instability that leads to the formation of an active region. Note that magnetic buoyancy instability is a threshold process \citep{Acheson79}, therefore a small increase of the  field gradient may be sufficient to trigger the emergence of the magnetic field 
already amplified by the stellar dynamo. 

The active region will rotate  with the orbital period of the planet,
although the finite Alfven travel time across the corona and the finite time scale of magnetic flux emergence may introduce an additional phase lag between the planet and the spot.

\end{document}